\documentclass{PoS}
\newcommand{\bea}{\begin{eqnarray}}
\newcommand{\eea}{\end{eqnarray}}
\newcommand{\beq}{\begin{equation}}
\newcommand{\eeq}{\end{equation}}
\newcommand{\nn}{\nonumber}

\usepackage[utf8x]{inputenc}
\usepackage{pgf}
\usepackage{amssymb,amsmath}
\usepackage{graphics}
\usepackage{graphicx}
\usepackage{epsfig}
\usepackage{pstricks}
\usepackage{tikz}

\def\bx{{\mathbf{x}}}
\def\by{{\mathbf{y}}}

\newcommand{\tr  }{\mathrm{Tr}}

\def\lsi{\raise0.3ex\hbox{$<$\kern-0.75em\raise-1.1ex\hbox{$\sim$}}}
\def\gsi{\raise0.3ex\hbox{$>$\kern-0.75em\raise-1.1ex\hbox{$\sim$}}}
\newcommand{\lsim}{\mathop{\lsi}}

\title{Onset Transition to Cold Nuclear Matter from Lattice QCD with 
Heavy Quarks to $\kappa^4$}

\ShortTitle{Onset Transition to Cold Nuclear Matter}

\author{\speaker{Jens Langelage}\\
        Institute for Theoretical Physics, ETH Z\"urich, CH-8093 Z\"urich,
Switzerland\\
        E-mail: \email{ljens@phys.ethz.ch}}

\author{Mathias Neuman\footnote{Speaker}, Owe Philipsen\\
        Institut f\"ur Theoretische Physik, Goethe-Universit\"at Frankfurt,\\
        60438 Frankfurt am Main, Germany\\
        E-mail: \email{neuman, philipsen@th.physik.uni-frankfurt.de}}

\abstract{
We present results of our ongoing studies of an effective three-dimensional theory
of thermal 
lattice QCD with heavy Wilson quarks. This is done by combined strong coupling and
hopping 
parameter expansions. The full quark determinant of four dimensional lattice QCD is
expanded in 
orders of the hopping parameter $\kappa$, the dimensional reduction is 
achieved by integrating over the spatial links. We present the 
calculation of the effective theory
through order $\kappa^nu^m$ with $n+m=4$.
This theory is then used to simulate heavy quarks near the cold and 
dense limit. For nonzero chemical potential the theory suffers from a sign problem,
wich is 
avoided by employing stochastical quantisation. 
Continuum extrapolated results for the onset of nuclear matter
are shown and the region of convergence of the effective theory is discussed.
}

\FullConference{31st International Symposium on Lattice Field Theory LATTICE 2013\\
                 July 29 - August 3, 2013\\
                 Mainz, Germany}

\begin{document}

\section{Introduction}

Thermal lattice QCD suffers from a severe sign problem when chemical potential is
nonvanishing. 
About a decade ago, several methods have been devised to circumvent this
obstacle 
(see e.g. \cite{deForcrand:2010ys} and references therein), but these 
are only valid for $\frac{\mu}{T}\lesssim1$. In order to go to higher chemical potentials, 
methods are required which at least potentially may solve this 
problem. Among these are Complex Langevin Dynamics (CLD) 
\cite{Aarts:2013bla,Aarts:2013uxa}, 
transformation of the degrees of freedom into so-called dual variables 
\cite{Gattringer:2012df,Delgado:2012tm}
and the formulation of the theory on a Lefschetz thimble \cite{Cristoforetti:2012su}.
But even if 
these approaches finally succeed in solving the sign problem, it will remain very
hard to study 
the region of cold and dense matter. This is because, in order to avoid 
the limiting artefact of saturation at finite lattice spacing, very fine
lattices are required for high density, which implies in turn very 
large temporal lattice extents near $T=0$. 
This motivates yet another approach, 
where we use strong coupling and hopping parameter expansions in order 
to simulate a 3d effective theory in a parameter regime where the 
sign problem is mild. See also 
\cite{Unger:2011it,Fromm:2011kq,Kawamoto:2005mq,Nakano:2010bg} for similar
approaches, where 
staggered fermions are being used. 
There the strong coupling series is much harder to 
compute, but with the advantage that the chiral regime can be studied. 

Here we show how to derive the $3$-dimensional effective theory by integrating out
the spatial 
degrees of freedom from the original $(3+1)d$ theory. This procedure has the 
additional 
benefit that the effective action can be formulated in terms of complex numbers
instead of 
group matrices. This allows us to simulate the effective 
theory quite fast and efficiently. 
Nevertheless, our approach also has some drawbacks, first and foremost that we do
not know 
the effective action in the full parameter regime. Our strategy is to expand 
the effective action around the static strong coupling limit, i.e. $\beta=\kappa=0$,
in a 
combined strong coupling and hopping parameter expansion. In 
previous works \cite{Fromm:2011qi,Fromm:2012eb} this has been shown to work rather
well and 
even allowing for continuum extrapolations in the heavy quark regime. Here, our
approach is
slightly 
adapted: In order to be able to describe physics near $T=0$ and large chemical
potentials, we 
expand in $\kappa$, but keep in each order the complete dependence on chemical
potential. 

\section{The effective action}

We start on a $(3+1)$-dimensional lattice with Wilsons gauge and
fermion 
actions, which after Grassmann integration may be written as
\begin{eqnarray}
Z=\int[dU_\mu]\det\left[Q\right]\exp\left[S_g\right]\;,\qquad
S_g=\frac{\beta}{2N_c}\sum_p\left[\tr\,   U_p+\tr\,   U_p^\dagger\right]\;,
\end{eqnarray}
where we defined the quark hopping matrix as
\begin{eqnarray}
Q^{ab}_{\alpha\beta,xy}=\delta^{ab}\delta_{\alpha\beta}\delta_{xy}-
\kappa\sum_{\nu=0}^3\left[e^{a\mu\delta_{\nu0}}(1+\gamma_\nu)_{\alpha\beta}U_\nu^{ab}(x)
\delta_{x,y-\hat{\nu}}+e^{-a\mu\delta_{\nu0}}(1-\gamma_\nu)_{\alpha\beta}U_{-\nu}^{ab}(x)
\delta_{x,y+\hat{\nu}}\right]
\;.\nonumber
\end{eqnarray}
Note that we consider only $N_f=1$ quark flavours in these first exploratory studies. 
The effective action is then defined by integrating out the spatial link variables
\begin{eqnarray}
e^{S_{\mathrm{eff}}}\equiv\int[dU_k]\det\left[Q\right]\exp\left[S_g\right]\;.
\label{eq_defeffth}
\end{eqnarray}
The crucial point of this approach is that the resulting effective theory does not
depend 
on the single temporal link variables, but only on their product along a temporal
axis, i.e. 
the Polyakov loops
\begin{eqnarray}
L_i\equiv\tr\,   W_i\equiv\prod_{\tau=1}^{N_\tau}U_0\left(\vec{x}_i;\tau\right)\;.
\end{eqnarray}
Our goal is now to expand eq.~(\ref{eq_defeffth}) in a combined strong coupling and
hopping 
parameter expansion. This introduces an infinite tower of effective interaction
terms, which 
 will be ordered according to their leading powers in $\beta, \kappa$. We will also make
sure that we 
 have the complete dependence on chemical potential in each order of the hopping
parameter  expansion, starting with the zeroth order, which is simply
pure gauge theory.

\subsection{Pure gauge theory}

In case of pure gauge theory it is advantageous to perform a character expansion
\begin{eqnarray}
\exp\left[\frac{\beta}{2N_c}\Big(\tr\,   U+\tr\,   U^\dagger\Big)\right]
=c_0(\beta)\left[1+\sum_{r\neq0}d_ra_r(\beta)\chi_r(U)\right]\;,
\end{eqnarray}
where the factor $c_0(\beta)$ can be neglected as it is independent of gauge links and
cancels in 
expectation values. In earlier publications
\cite{Fromm:2011qi,Langelage:2010yr,Langelage:2010nj}, we have shown how to compute
the effective 
gauge theory up to rather high orders in the fundamental character expansion
coefficient 
$u(\beta)\equiv a_f(\beta)$. In leading order we have a chain of $N_\tau$ fundamental 
plaquettes winding around the temporal direction and closing via periodic boundary
conditions.        
It reads
\begin{eqnarray}
e^{S_{\mathrm{eff}}^{(1)}}=\lambda(u,N_\tau)\sum_{<ij>}\left(L_iL_j^\ast+L_i^\ast
L_j\right)\;,
\qquad\lambda(u,N_\tau)=u^{N_\tau}\Big[1+\ldots\Big]\;,
\label{eq_seffgauge}
\end{eqnarray}
where higher order corrections of $\lambda(u,N_\tau)$ as well as a discussion of
higher order 
interaction terms can be found in \cite{Langelage:2010yr}. In the leading order
expression of eq.~(\ref{eq_seffgauge}) we already see that 
$\lambda(u,N_\tau)$ is suppressed for large $N_\tau$, since $u<1$, see also
\cite{Fromm:2011qi} 
for a further discussion of this aspect.

\subsection{Static quark determinant}

Let us now expand the quark determinant in a hopping expansion. In order to keep the
complete 
dependence on chemical potential, we split the quark matrix according to
\begin{eqnarray}
Q=1-T-S=1-T^+-T^--S^+-S^-\;,
\end{eqnarray}
in positive and negative temporal and spatial parts. The static determinant is
then given 
by neglecting the spatial parts.
We define and compute the static determinant to be
\begin{eqnarray}
\det[Q_{\mathrm{stat}}] &=& \det[1-T] = \det[1-T^+ - T^-] \nn \\
&=& \det \Big[1-\kappa e^{a\mu}(1+\gamma_0)U_0 \delta_{x,y-\hat{0}} - 
\kappa e^{-a\mu}(1-\gamma_0)U^{\dagger}_0 \delta_{x,y+\hat{0}}\Big]
\end{eqnarray}
with propagation in the temporal direction only. Perfoming the space and spin
determinant we 
get
\begin{eqnarray}
\det[Q_{\mathrm{stat}}]&=& \prod_{\vec{x}} 
\det \Big[1+(2 \kappa e^{a \mu})^{N_{\tau}}W_{\vec{x}}\Big]^2
\det \Big[1+(2 \kappa e^{-a \mu})^{N_{\tau}}W^{\dagger}_{\vec{x}}\Big]^2\;.
\label{q_static}
\end{eqnarray}
A well-known
relation valid 
for $SU(3)$ then 
allows us to reformulate this in terms of traced Polyakov loops
\begin{eqnarray}
\det[Q _{\mathrm{stat}}]&=& \prod_{\vec{x}} 
\left[1 + c L_{\vec{x}} + c^2 L^{\dagger}_{\vec{x}}+c^3\right]^2
\left[1 + \bar{c} L^{\dagger}_{\vec{x}} + \bar{c}^2 L_{\vec{x}}+\bar{c}^3\right]^2,
\label{eq_qsim}
\end{eqnarray}
with $c=c(\mu)=\left(2\kappa e^{a\mu}\right)^{N_\tau}=\bar{c}(-\mu)$ in the strong
coupling limit.

\subsection{Kinetic quark determinant}

In order to compute a systematic hopping expansion, we define the kinetic quark
determinant as 
follows
\begin{eqnarray}
\det[Q]&\equiv&\det[Q_{\mathrm{stat}}][Q_{\mathrm{kin}}]\;,\nonumber\\
\det[Q_{\mathrm{kin}}]&=&[1-(1-T)^{-1}(S^++S^-)]\equiv\det[1-P-M]=\exp\left[\tr\,  \ln
(1-P-M)\right]\;,
\label{eq_detqkin}
\end{eqnarray}
which we then split into parts describing quarks moving in positive and negative
spatial 
directions, $P=\sum_kP_k$ and $M=\sum_kM_k$. The reason for this is that the trace
occurring 
in eq.~(\ref{eq_detqkin}) is also a trace in coordinate space. This means
that only closed loops contribute and hence
we need the same number of $P$s and $M$s in the expansion of the logarithm.
Through $\mathcal{O}\left(\kappa^4\right)$ we have
\begin{eqnarray}
\det[Q_{\mathrm{kin}}]&=&\exp\left[-\tr\,   PM-\tr\,   PPMM-
\frac{1}{2}\tr\,   PMPM\right]\left[1+\mathcal{O}(\kappa^6)\right]\nonumber\\
&=&\left[1-\tr\,   PM-\tr\,   PPMM-
\frac{1}{2}\tr\,   PMPM+\frac{1}{2}\left(\tr\,   PM\right)^2\right]
\left[1+\mathcal{O}(\kappa^6)\right]\;.
\label{eq_detqkin2}
\end{eqnarray}
The next step is now to consider the different directions in $P$ and 
$M$ and to neglect the vanishing contributions, i.e. those which have e.g. a $P_k$
but no $M_k$
\begin{eqnarray}
\sum_{kl}\tr\,   P_kM_l&=&\sum_k\tr\,   P_kM_k\;,\label{eq_qdet2}\\
\sum_{klmn}\tr\,  P_kP_lM_mM_n&=&\sum_k\tr\,  P_kP_kM_kM_k+
\sum_{k\neq l}\tr\,  P_kP_lM_kM_l+\sum_{k\neq
l}\tr\,  P_kP_lM_lM_k\label{eq_ppmm}\;,\\
 \frac12 \sum_{klmn}\tr\,  P_kM_lP_mM_n&=& \frac12 \sum_k\tr\,  P_kM_kP_kM_k+
 \frac12 \sum_{k\neq l}\tr\,  P_kM_kP_lM_l+ \frac12 \sum_{k\neq
l}\tr\,  P_kM_lP_lM_k\label{eq_pmpm}\;,\\
\frac12 \sum_{klmn}\tr\,  P_kM_l\tr\,  P_mM_n&=&\frac12 \sum_{k,
l}\tr\,  P_kM_k\tr\,  P_lM_l\label{eq_trpm2}\;.
\end{eqnarray}
Having these expressions, the final ingredient is to compute the static quark
propagator 
$(1-T)^{-1}$, appearing in eq.~(\ref{eq_detqkin}).

\subsection{Static quark propagator}

Since $(1+\gamma_\mu)(1-\gamma_\mu)=0$, hops in forward and backward time
direction 
do not mix and
 the full static quark propagator is given by
\begin{eqnarray*}
(Q_{\mathrm{stat}})^{-1}=(Q^+_{\mathrm{stat}})^{-1}
+(Q^-_{\mathrm{stat}})^{-1}-1\;.
\end{eqnarray*}
In order to compute the positive static quark propagator, we 
use the series expansion
\begin{eqnarray*}
(Q^+_{\mathrm{stat}})^{-1}=\left(1-T^+\right)^{-1}=\sum_{n=0}^\infty
(T^+)^n\;,
\end{eqnarray*}
where convergence is only guaranteed for $z\equiv2\kappa e^{a\mu}<1$. 
The inverse is given by
\begin{eqnarray*}
(Q^+_{\mathrm{stat}})^{-1}_{\tau_1\tau_2}&=&\delta_{\tau_1\tau_2}\left(1-qcW\right)+qz^{\tau_2-\tau_1}W(\tau_1,\tau_2)\Big[\Theta(\tau_2-\tau_1)-z^{N_\tau}
\Theta(\tau_1-\tau_2)\Big]\;,
\end{eqnarray*}
where 
\begin{eqnarray*}
q\equiv\frac{1}{2}(1+\gamma_0)\left(1+cW\right)^{-1}\;,
\end{eqnarray*}
and $W(\tau_1,\tau_2)$ is a temporal Wilson line from $\tau_1$ to $\tau_2$. If 
$\tau_1=\tau_2$, i.e. the Wilson loop winds around the lattice, we have the usual
(untraced) 
Polyakov loop $W(\tau_1,\tau_1)=W$. Although we have derived this expression with a
geometric series, which converges only for 
$c<1$, it can be shown that this is indeed the inverse for all values of $c$,
e.g. by 
evaluating that $(Q^+_{\mathrm{stat}})^{-1}(Q^+_{\mathrm{stat}})=1$.

The contribution in negative time direction 
$(Q^-_{\mathrm{stat}})^{-1}_{\tau_1\tau_2}$ can then be obtained from 
$(Q^+_{\mathrm{stat}})^{-1}_{\tau_1\tau_2}$ by the following replacements
\begin{eqnarray*}
\tau_1\leftrightarrow\tau_2\;,\qquad
W(\tau_1,\tau_2)\leftrightarrow W^\dagger(\tau_1,\tau_2)\;,\qquad
\mu\leftrightarrow-\mu\;,
\end{eqnarray*}
and reads
\begin{eqnarray*}
(Q^-_{\mathrm{stat}})^{-1}_{\tau_1\tau_2}&=&\delta_{\tau_1\tau_2}\left(1-\bar{q}\bar{c}W^\dagger
\right)+\bar{q}\bar{z}^{\tau_1-\tau_2}W^\dagger(\tau_1,\tau_2)\Big[\Theta(\tau_1-\tau_2)-\bar{z}^{N_\tau}
\Theta(\tau_2-\tau_1)\Big]\;,\\
\bar{q}&=&\frac{1}{2}(1-\gamma_0)\left(1+\bar{c}W^\dagger\right)^{-1}\;,\qquad
\bar{z}=2\kappa e^{-a\mu}\;.
\end{eqnarray*}
Finally we split the temporal quark propagator in spin space as well as in
propagation in positive 
and negative temporal direction according to
\begin{eqnarray}
\label{eq_qstat}
\left(Q_{\mathrm{stat}}\right)^{-1}&=& A + \gamma_0 B = A^++\gamma_0B^+ + A^--\gamma_0 B^-\;,\\
A^+_{xy}&=&\frac12 \left[1-\frac{cW}{1+cW}\right]\delta_{xy}
+\frac{1}{2}z^{\tau_y-\tau_x}\frac{W(\tau_x,\tau_y)}{1+cW}\bigg[\Theta(\tau_y-\tau_x)-c
\Theta(\tau_x-\tau_y)\bigg]\delta_{\bx\by}\;,\nonumber\\
B^+_{xy}&=&-\frac{1}{2}\frac{cW}{1+cW}\delta_{xy}
+\frac{1}{2}z^{\tau_y-\tau_x}\frac{W(\tau_x,\tau_y)}{1+cW}\bigg[\Theta(\tau_y-\tau_x)-c
\Theta(\tau_x-\tau_y)\bigg]\delta_{\bx\by}\;,\nonumber\\
A^-_{xy}&=&\frac12 \left[1-1\frac{\bar{c}W^\dagger}{1+\bar{c}W^\dagger}\right]\delta_{xy}
+\frac{1}{2}\bar{z}^{\tau_x-\tau_y}\frac{W^\dagger(\tau_x,\tau_y)}{1+\bar{c}W^\dagger}\bigg[\Theta(\tau_x-\tau_y)-\bar{c}
\Theta(\tau_y-\tau_x)\bigg]\delta_{\bx\by}\;,\nonumber\\
B^-_{xy}&=&-\frac{1}{2}\frac{\bar{c}W^\dagger}{1+\bar{c}W^\dagger}\delta_{xy}
+\frac{1}{2}\bar{z}^{\tau_x-\tau_y}\frac{W^\dagger(\tau_x,\tau_y)}{1+\bar{c}W^\dagger}\bigg[\Theta(\tau_x-\tau_y)-\bar{c}
\Theta(\tau_y-\tau_x)\bigg]\delta_{\bx\by}\;,\nonumber
\end{eqnarray}
Due to the length of these terms, we will formulate our results usually in $A$ and
$B$ for brevity.

\subsection{The leading correction terms}

Now it is time to perform the group integrations. Let us for notational convenience
define the following quantities
\begin{eqnarray}
\int[dU_k]\det[Q_{\mathrm{kin}}]\equiv1 + \sum_{n,m} \Delta^{(n,m)}\;,
\end{eqnarray}
where $n$ denotes the order in the hopping parameter $\kappa$ and $m$ specifies the mth term appearing 
in
eqs~(\ref{eq_qdet2}-\ref{eq_trpm2}).

\subsubsection{$\tr\,  PM$:}

From eq.~(\ref{eq_qdet2}) and after a few steps of algebra the correction of 
$\mathcal{O}(\kappa^2)$ is given by
\begin{eqnarray*}
\Delta^{(2,1)}&\equiv&\int[dU_k]\sum_{i}\tr\,  P_iM_i=
\sum_i\int[dU_k]\tr\,  \Big[(Q_{\mathrm{stat}}^+)^{-1}S^+_i(Q_{\mathrm{stat}}^+)^{-1}
S^-_i\Big]\nonumber\\
&=&-\frac{8 \kappa^2}{N_c}\sum_{u,i} \tr\,   B_{u,u}
\tr\,B_{u +\hat{\imath},u+\hat{\imath}} \nn \\
&=& -\frac{2 \kappa^2 N_{\tau}}{N_c} \sum_{\vec{x},i} \Bigg[\bigg(\tr\,  
\frac{c  W_{\vec{x}}}{1 + c W_{\vec{x}}} - \tr\, \frac{\bar{c}
W^{\dagger}_{\vec{x}}}{1 + \bar{c} 
W^{\dagger}_{\vec{x}}} \bigg)\bigg(
\tr\,   
\frac{c W_{\vec{x}+\hat{\imath}}}{1 + c W_{\vec{x}+\hat{\imath}}}
- \tr\,  \frac{\bar{c}  
W^{\dagger}_{\vec{x}+\hat{\imath}}}{1 + \bar{c}  W^{\dagger}_{\vec{x}+\hat{\imath}}}
\bigg) 
\Bigg]
\end{eqnarray*}
where we have used the expressions eq.~(\ref{eq_qstat}) for $B$ and
evaluated the trace over 
spin and coordinate space. The group integrations 
have been computed via
\begin{eqnarray}
\int dU U_{ij}U^\dagger_{kl}=\frac{1}{3}\delta_{il}\delta_{jk}\;.
\end{eqnarray}
Note that this enforces the spatial link variables to be at the same 
temporal location and yields a factor $N_\tau$ rather than $N_\tau^2$ 
from the two temporal traces. From now on we will skip the last step, where one 
has to insert the definitions of $A$ and $B$ and perform the temporal sums.

\subsubsection{$\tr\,  PPMM$:}

Here we have three contributions according to eq.~(\ref{eq_ppmm}), which after group
integration and tracing read
\begin{eqnarray*}
\Delta^{(4,1)}&=&-\frac{32 \kappa^4}{N_c^2} \sum_{u,v,i} \tr\,
B_{u,u}\tr\,A_{u + \hat{\imath}, v + \hat{\imath}} A_{v + \hat{\imath}, u + \hat{\imath}} \tr\,B_{u + 2 \hat{\imath}, u + 2 \hat{\imath}}\;,\\ 
\Delta^{(4,2)}&=&\mathcal{O}(\kappa^4u)\;,\\
\Delta^{(4,3)}&=&-\frac{16 \kappa^4}{N_c^2}
\sum_{u,v,i \neq j}
\tr\,B_{u - \hat{\imath},u - \hat{\imath}}
\Big[\tr\,A_{u,v}A_{v,u}+\tr\,B_{u,v}B_{v,u}\Big]
\tr\,B_{u + \hat{\jmath},u + \hat{\jmath}}\;,
\end{eqnarray*}
where in case of $\Delta^{(4,2)}$ one has to leave the strong coupling limit and
include an
additional gauge plaquette due to otherwise vanishing group integration.

\subsubsection{$\tr\,  PMPM$:}

The contributions in this term read following eq.~(\ref{eq_pmpm})
\begin{eqnarray*}
\Delta^{(4,4)}&=&
-\frac{16\kappa^4}{N_c^2}\sum_{u \neq v, i}\left[\tr\,
B_{u,v}B_{v,u}
\Big(\tr\, B_{u+\hat{\imath},u+\hat{\imath}}\Big)^2 +
\Big(\tr\,B_{u,u}\Big)^2 
\tr\,B_{u+\hat{\imath},v+\hat{\imath}}B_{v+\hat{\imath},u+\hat{\imath}} \right]\nn\\
&&-\frac{16\kappa^4}{(N_c^2 -
1)}\sum_{u, i}\bigg\lbrace\tr\,B_{u,u}B_{u,u}
\Big(\tr\, B_{u+\hat{\imath},u+\hat{\imath}}\Big)^2 +
\Big(\tr\,B_{u,u}\Big)^2 
\tr\,B_{u+\hat{\imath},u+\hat{\imath}}B_{u+\hat{\imath},u+\hat{\imath}}\nn\\
&&-\frac{1}{N_c}\left[\tr\,B_{u,u}B_{u,u} \tr\,B_{u +\hat{\imath}, u +\hat{\imath}}B_{u +\hat{\imath}, u +\hat{\imath}}
 + \Big(\tr\, B_{u,u}\Big)^2 \Big(\tr\,
B_{u+\hat{\imath},u+\hat{\imath}}\Big)^2\right]\bigg\rbrace\;,\label{eq_det44}
\\
\Delta^{(4,5)}&=&-\frac{8 \kappa^4}{N_c^2}\sum_{u,v,i \neq j}
\tr\,B_{u-\hat{\imath},u-\hat{\imath}}
\Big[\tr\,A_{u,v}A_{v,u}+\tr\,B_{u,v}B_{v,u}\Big]
\tr\, B_{u+\hat{\jmath},u+\hat{\jmath}})\;,\\
\Delta^{(4,6)}&=&-\frac{8 \kappa^4}{N_c^2}\sum_{u,v,i \neq j}
\tr\,B_{u-\hat{\imath},u-\hat{\imath}}
\Big[\tr\,A_{u,v}A_{v,u}+\tr\,B_{u,v}B_{v,u}\Big]
\tr\, B_{u-\hat{\jmath},u-\hat{\jmath}})\;,\\.
\end{eqnarray*}
In the calculation of $\Delta^{(4,4)}$ it may happen that there is a spatial
link which is occupied by four matrices and we need the group integral (see e.g.
\cite{Creutz:1978ub})
\begin{eqnarray}
\int
dU\,U_{i_1j_1}U_{i_2j_2}U^\dagger_{k_1l_1}U^\dagger_{k_2l_2}&=&\frac{1}{N_c^2-1}\Big[\delta_{i_1l_1}\delta_{i_2l_2}\delta_{j_1k_1}\delta_{j_2k_2}+\delta_{i_1l_2}\delta_{i_1l_2}\delta_{j_1k_2}\delta_{j_2k_1}\Big]\nonumber\\
&-&\frac{1}{N_c(N_c^2-1)}\Big[\delta_{i_1l_2}\delta_{i_2l_1}\delta_{j_1k_1}\delta_{j_2k_2}+\delta_{i_1l_1}\delta_{i_2l_2}\delta_{j_1k_2}\delta_{j_2k_1}\Big]\;.
\end{eqnarray}

\subsubsection{$(\tr\,  PM)^2$:}

Here we have to consider two different possibilities: The two nearest-neighbour 
contributions may share $0$, $1$ or $2$ sites, where the first two lead to
the same result. Their contribution is given by
\begin{eqnarray*}
\Delta^{(4,7)}=\frac{32\kappa^4}{N_c^2}\sum_{u,v, (u+\hat{\imath} \neq v +\hat{\jmath} )}
 \tr\,   B_{u,u} \tr\,B_{u+\hat{\imath},u+\hat{\imath}}\tr\,  
B_{v,v} \tr\,B_{v+\hat{\jmath},v+\hat{\jmath}}\;.
\end{eqnarray*}
The other possibility is that $(\vec{u},i)=(\vec{v},j)$, i.e. we have only two sites
involved and get
\begin{eqnarray*}
\Delta^{(4,8)}&=&\frac{32\kappa^4}{N_c^2}\sum_{u \neq v, i}\left[ \Big(\tr\, B_{u,u}\Big)^2 \Big(\tr\,
B_{v+\hat{\imath},v+\hat{\imath}}\Big)^2
+\tr\,B_{u,v}B_{v,u} \tr\,B_{u+\hat{\imath},v+\hat{\imath}}B_{v+\hat{\imath},u+\hat{\imath}}\right]\nn\\
&&+\frac{32\kappa^4}{N_c^2-1}\sum_{u,i}\Bigg\lbrace\Big(\tr\,
B_{u,u}(\vec{x})\Big)^2 \Big(\tr\,
B_{u+\hat{\imath},u+\hat{\imath}}\Big)^2+\tr\,B_{u,u}B_{u,u}\tr\,B_{u+\hat{\imath},u+\hat{\imath}}
B_{u+\hat{\imath},u+\hat{\imath}}\nn\\
&&-\frac{1}{N_c}\bigg[\tr\,B_{u,u}B_{u,u}
\Big(\tr\, B_{u+\hat{\imath},u+\hat{\imath}}\Big)^2 +
\Big(\tr\,B_{u,u}\Big)^2 
\tr\,B_{u+\hat{\imath},u+\hat{\imath}}B_{u+\hat{\imath},u+\hat{\imath}}
\bigg]\Bigg\rbrace\;.
\end{eqnarray*}

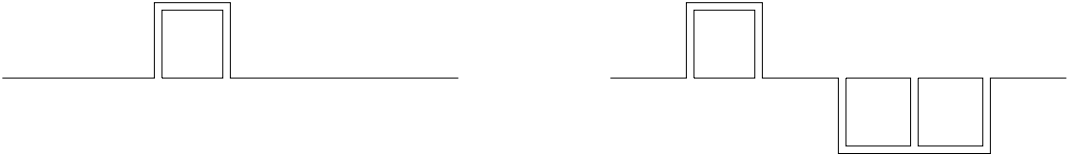
\begin{figure}
\scalebox{0.5}{
\begin{tikzpicture}
\draw(16,0) -- (20,0) -- (20,2) -- (22,2) -- (22,0) -- (28,0);
\draw(20.2,0) -- (20.2,1.8) -- (21.8,1.8) -- (21.8,0) -- (20.2,0);
\draw(32,0) -- (34,0) -- (34,2) -- (36,2) -- (36,0) -- (38,0) -- (38,-2) -- (42,-2)
-- (42,0) -- 
(44,0);
\draw(34.2,0) -- (34.2,1.8) -- (35.8,1.8) -- (35.8,0) -- (34.2,0);
\draw(38.2,0) -- (38.2,-1.8) -- (39.9,-1.8) -- (39.9,0) -- (38.2,0);
\draw(40.1,0) -- (40.1,-1.8) -- (41.8,-1.8) -- (41.8,0) -- (40.1,0);
\end{tikzpicture}
}
\caption{Finite gauge coupling corrections to the Polyakov line. After spatial link
integration 
these graphs give rise to terms $\sim \tr W$.}
\label{fig_pl}
\end{figure}

\subsection{Resummations}

In order to include as many terms as possible and improve convergence we perform a
resummation. The 
contributions of $\tr\,PM$ and parts of $\frac{1}{2}\tr\,PM \tr\,PM$ can be written
as an 
exponential:
\begin{eqnarray}
\exp\left[\Delta^{(2,1)}\right]&=&1- \frac{8 \kappa^2}{N_c} \sum_{u,i} \tr\,  
B_{u,u} \tr\,B_{u+\hat{\imath},u+\hat{\imath}} \nn\\
&&+\frac{32 \kappa^4}{N^2_c}\sum_{u,v,i,j}
  \tr\,   B_{u,u} \tr\,B_{u+\hat{\imath},u+\hat{\imath}}\tr\,  
B_{v,v} \tr\,B_{v+\hat{\jmath},v+\hat{\jmath}} + 
\ldots
\end{eqnarray}
Inspection of higher order terms indicates that this should always be possible.
Therefore we may 
write
\begin{eqnarray}
\int[dU_k]\det[Q_{\mathrm{kin}}] = e^{\sum_{n,m} \Delta^{(n,m)}}\;,
\label{eq_exp}
\end{eqnarray}
where $\Delta^{(4,7)}$ and the parts of $\Delta^{(4,8)}\sim \big(\Delta^{(2,1)}\big)^2$ are to be excluded to avoid double counting.

\subsection{Leading gauge corrections to the strong coupling limit}

Leaving the strong coupling limit, i.e. $\beta \neq 0$, gauge plaquettes have to be
included. This 
makes the effective coupling constants depend on the gauge coupling: 
$h_i(\kappa)\rightarrow h_i(\kappa,u)$. A somewhat special role plays the
single Polyakov line coupling 
$c$ introduced in eq.~(\ref{eq_qsim}), since it also enters in the static propagator 
eq.~(\ref{eq_qstat}). Hence we may further resum terms by replacing the static
version $c$ with 
$h_1$, which is defined to include gauge interactions. 
The leading gauge corrections are of order $N_{\tau} \kappa^2 u$ coming from
attaching plaquettes 
to the Wilson line, cf. fig.~\ref{fig_pl}
\begin{eqnarray}
c \rightarrow h_1 = c \ \Big[1+6\kappa^2 N_{\tau} u + \mathcal{O}(\kappa^2 u^5) \Big].
\end{eqnarray}
This can also be exponentiated by summing over multiple attached plaquettes at
different locations
\begin{eqnarray}
h_1&=& c \exp \left[6 \kappa^2 N_{\tau} \frac{u - u^{N_{\tau}}}{1-u} \right]=
\exp\left[N_\tau\left(a\mu+\ln2\kappa+6 \kappa^2 \frac{u -
u^{N_{\tau}}}{1-u}\right)\right]\;,
\end{eqnarray}
and we see that in this way the Polyakov line receives mass corrections due to
interactions.
Note that this generates overcounting in higher orders, but in our opinion the 
resummation effects of this procedure more than compensates for this additional
care. Let us finally 
also give a correction for the coefficient in $\Delta^{(2,1)}$
\begin{eqnarray}
\frac{2 \kappa^2 N_{\tau}}{N_c} \rightarrow \frac{2\kappa^2N_\tau}{N_c}\left[1+2\frac{u-u^{N_\tau}}{1-u}+\ldots\right]\;.
\end{eqnarray}
With this the effective threedimensional theory we are going to simulate is finally
given by
\begin{eqnarray}
Z_{\mathrm{eff}}=\int[dW]\det[Q_{\mathrm{stat}}]\exp\left[\Delta^{(2,1)}+
\sum_{m} \Delta^{(4,m)}\right]\;,
\end{eqnarray}
where the sum over $m$ is restricted in the same way as in eq.~(\ref{eq_exp}).

\section{Simulation of the effective theory by complex Langevin}
\begin{figure}[t]
\centerline{
\includegraphics[width=0.5\textwidth]{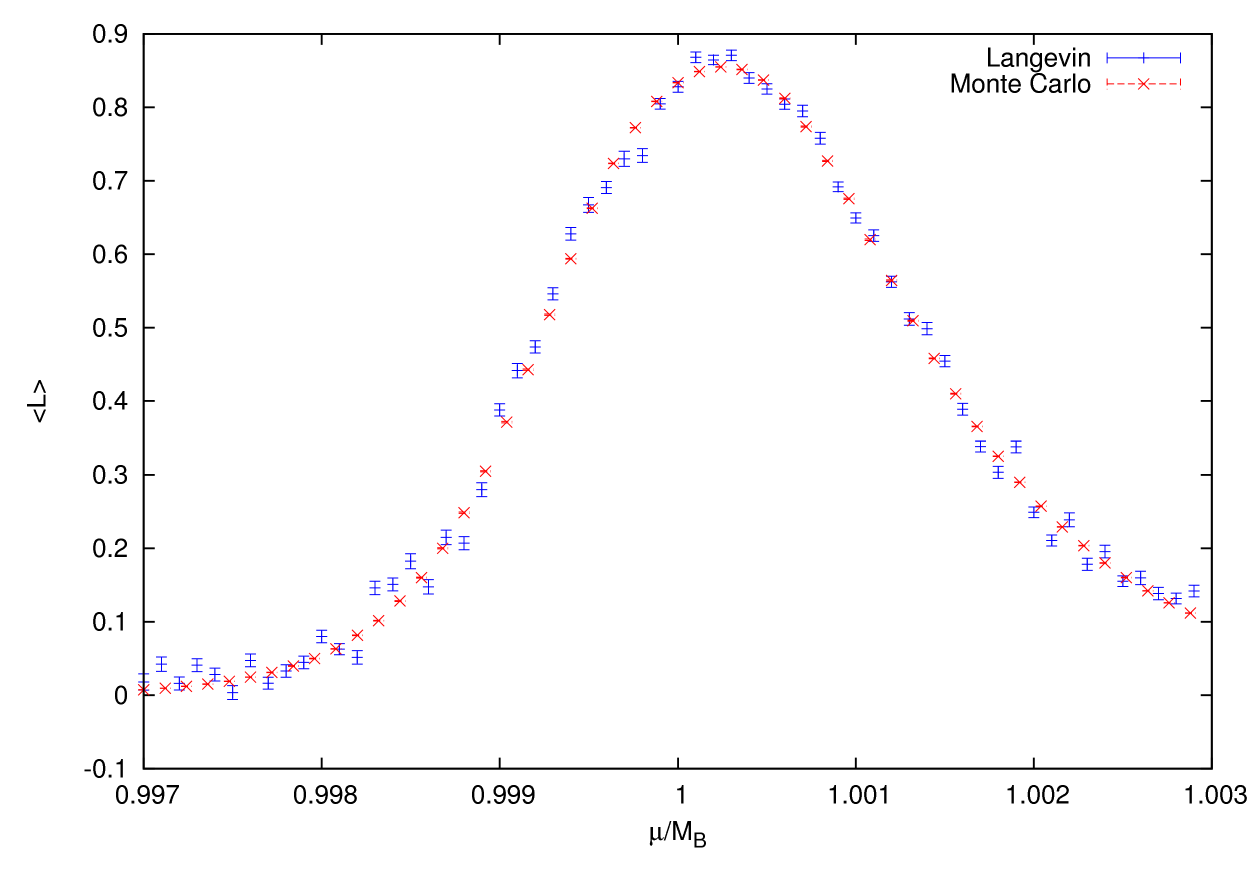}
\includegraphics[width=0.5\textwidth]{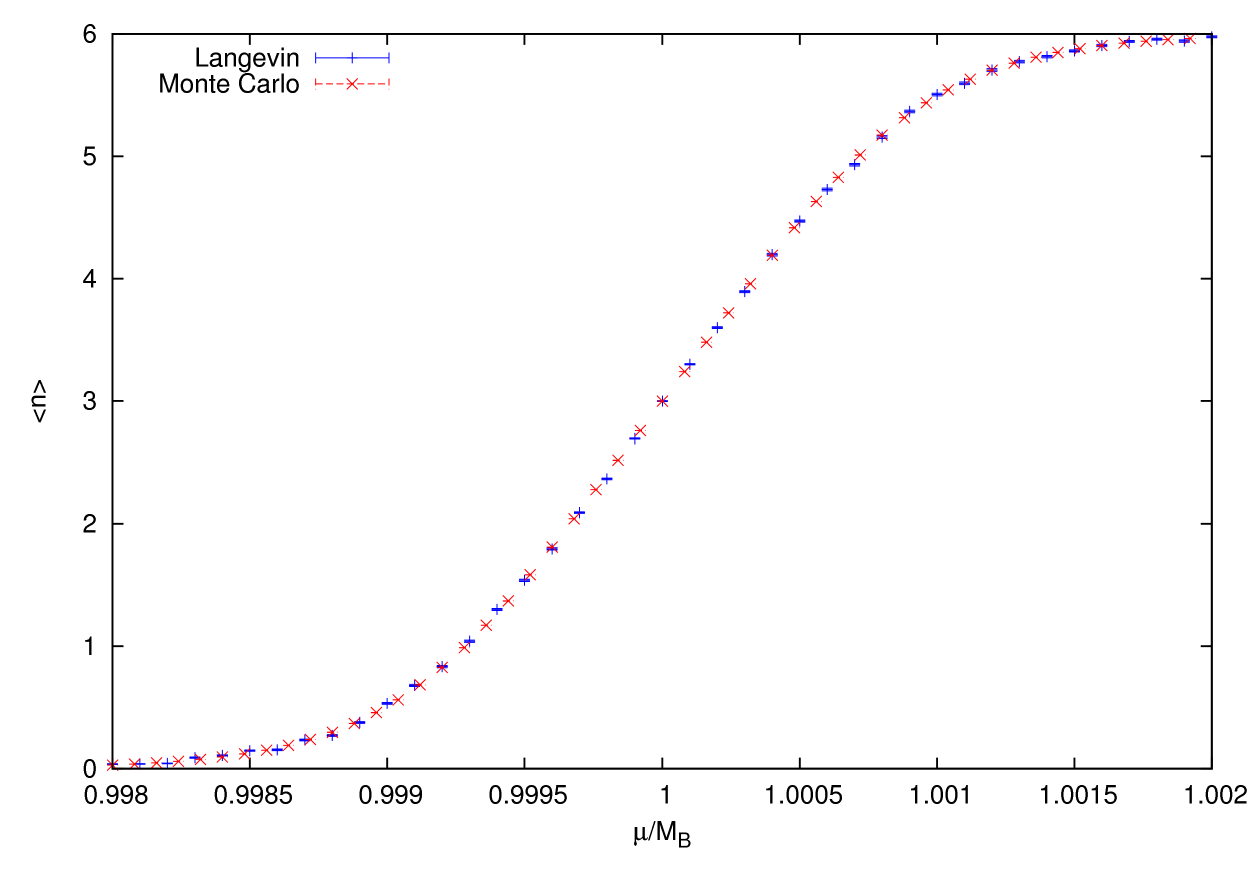}
}
\caption[]{Comparison between Langevin and Monte Carlo data at $\kappa=0.01$ and
$N_{\tau}=200$ in the strong coupling limit.}
\label{fig:cfMC}
\end{figure}

\begin{figure}[t]
\centerline{
\includegraphics[width=0.5\textwidth]{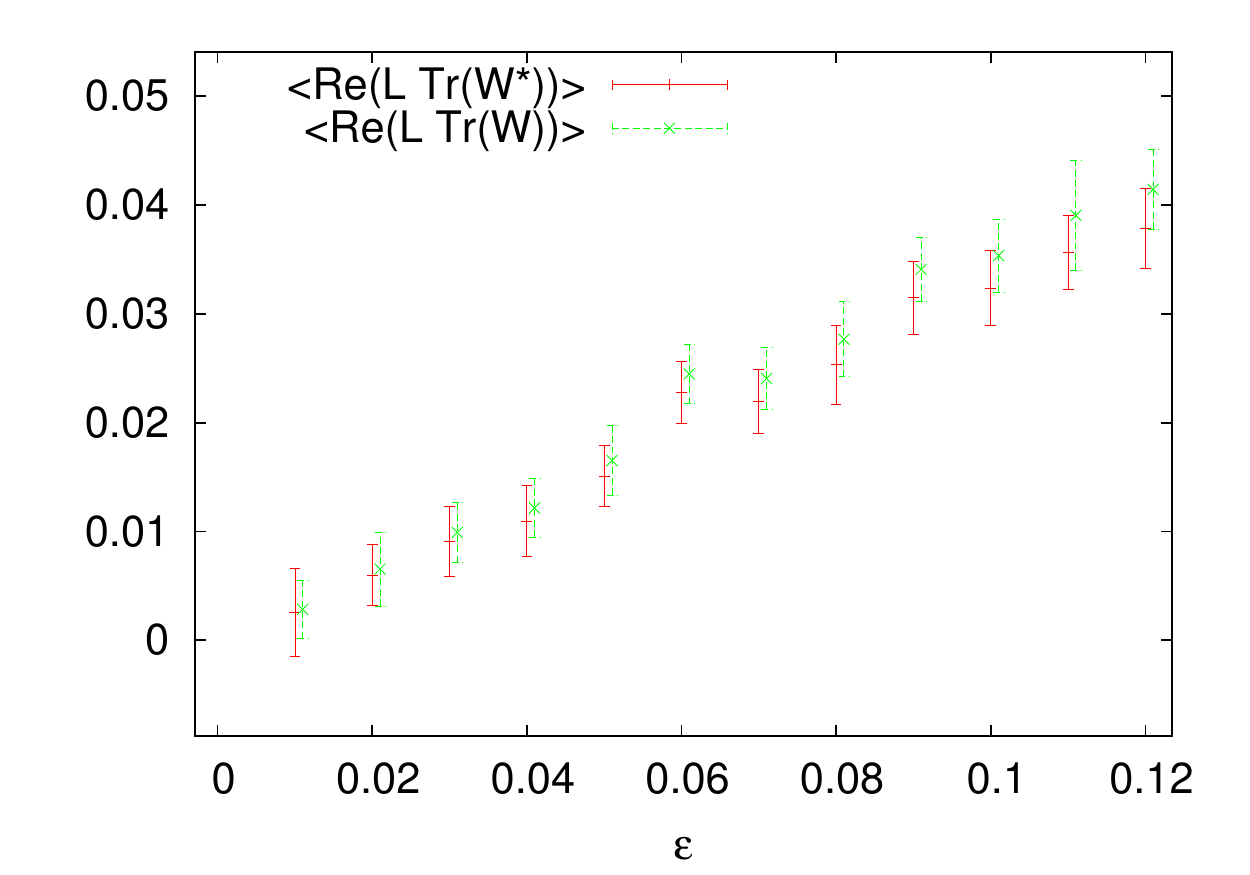}
\includegraphics[width=0.5\textwidth]{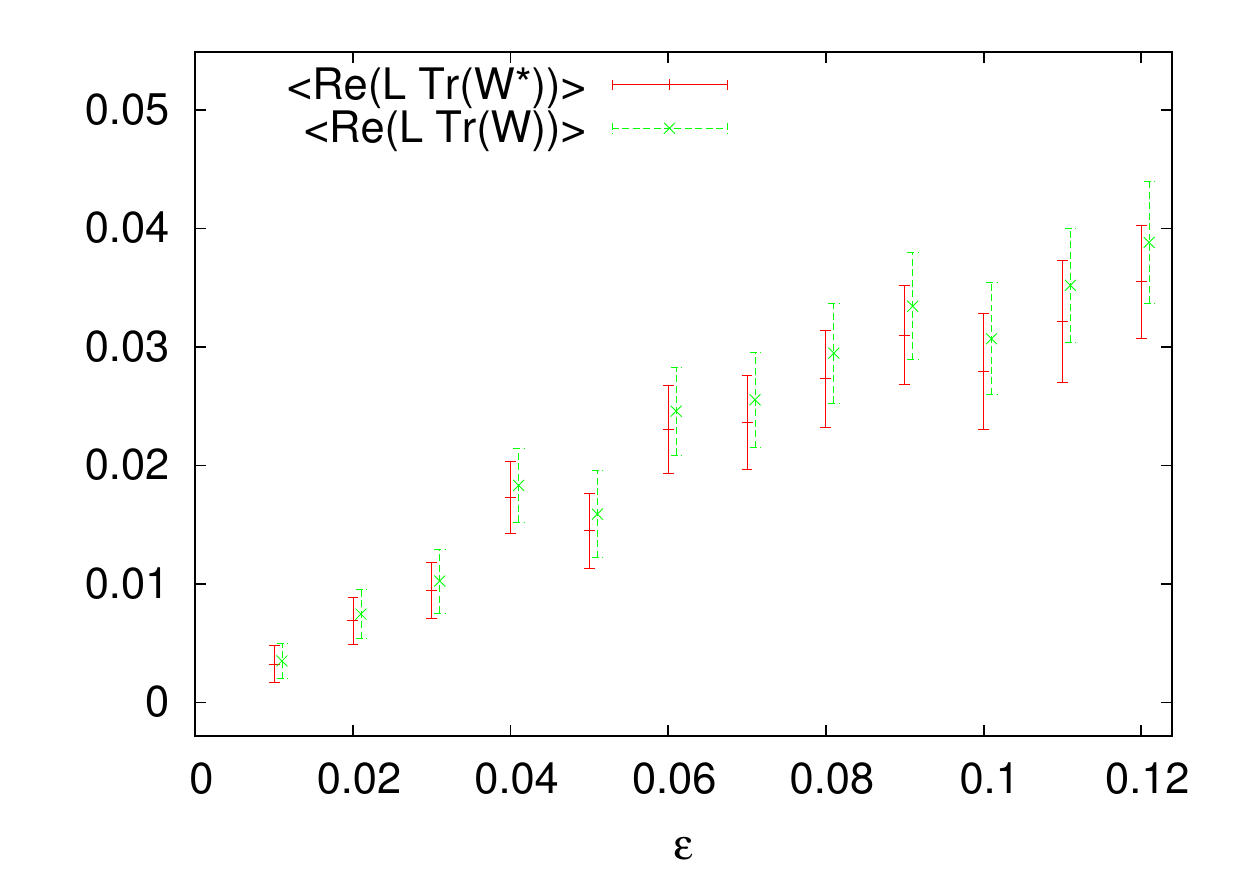}
}
\caption[]{Test of the convergence criterion for complex Langevin in the effective
theory to order
$\kappa^2$ (left) and $\kappa^4$ (right) for $\frac{\kappa^2 N_{\tau}}{N_c} = 0.01 $ and $\beta = 5.7$. $L$ refers to the operator in (\ref{langevan operator}).}
\label{fig:convcrit}

\end{figure}

\begin{figure}[t]
\hspace*{1cm}
\scalebox{0.85}{
\centerline{
\includegraphics[width=0.4\textwidth]{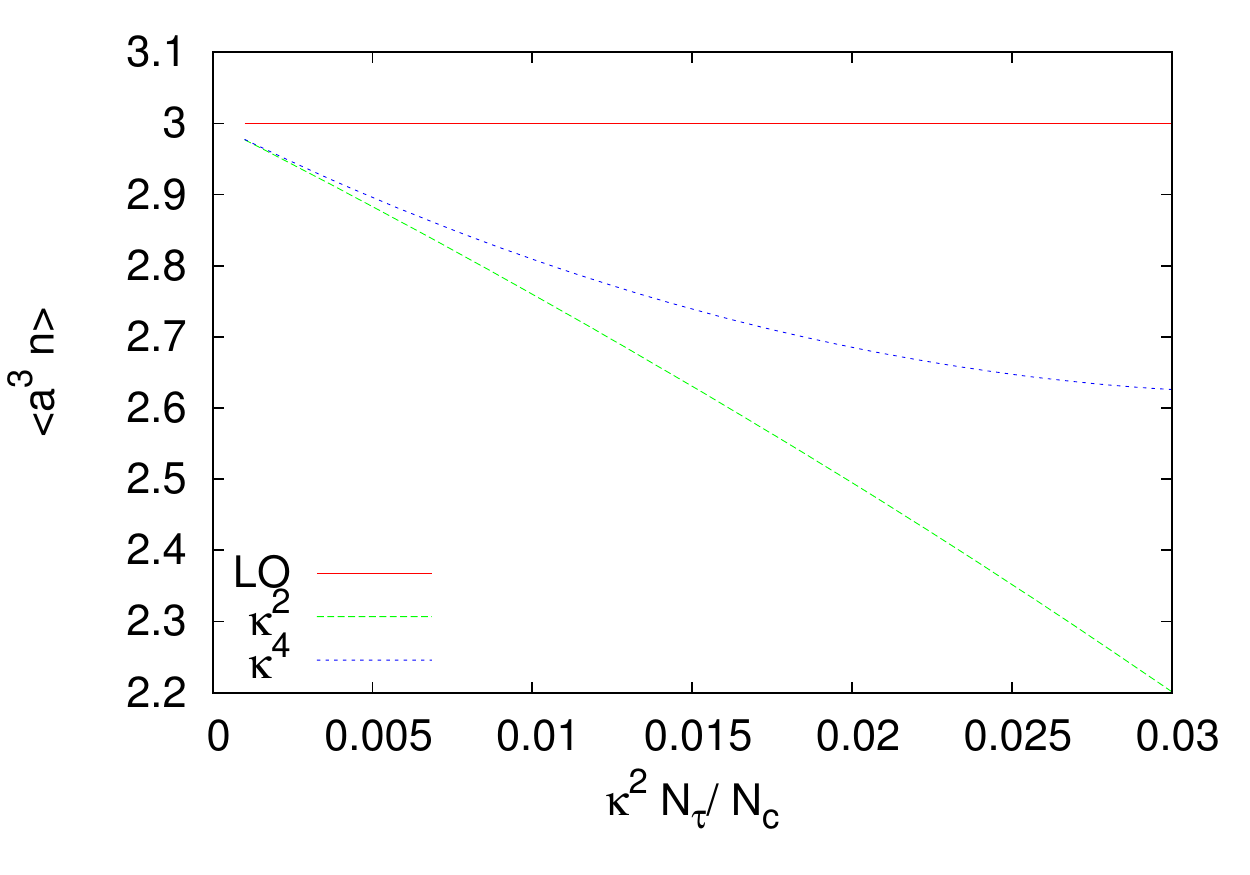}
\includegraphics[width=0.4\textwidth]{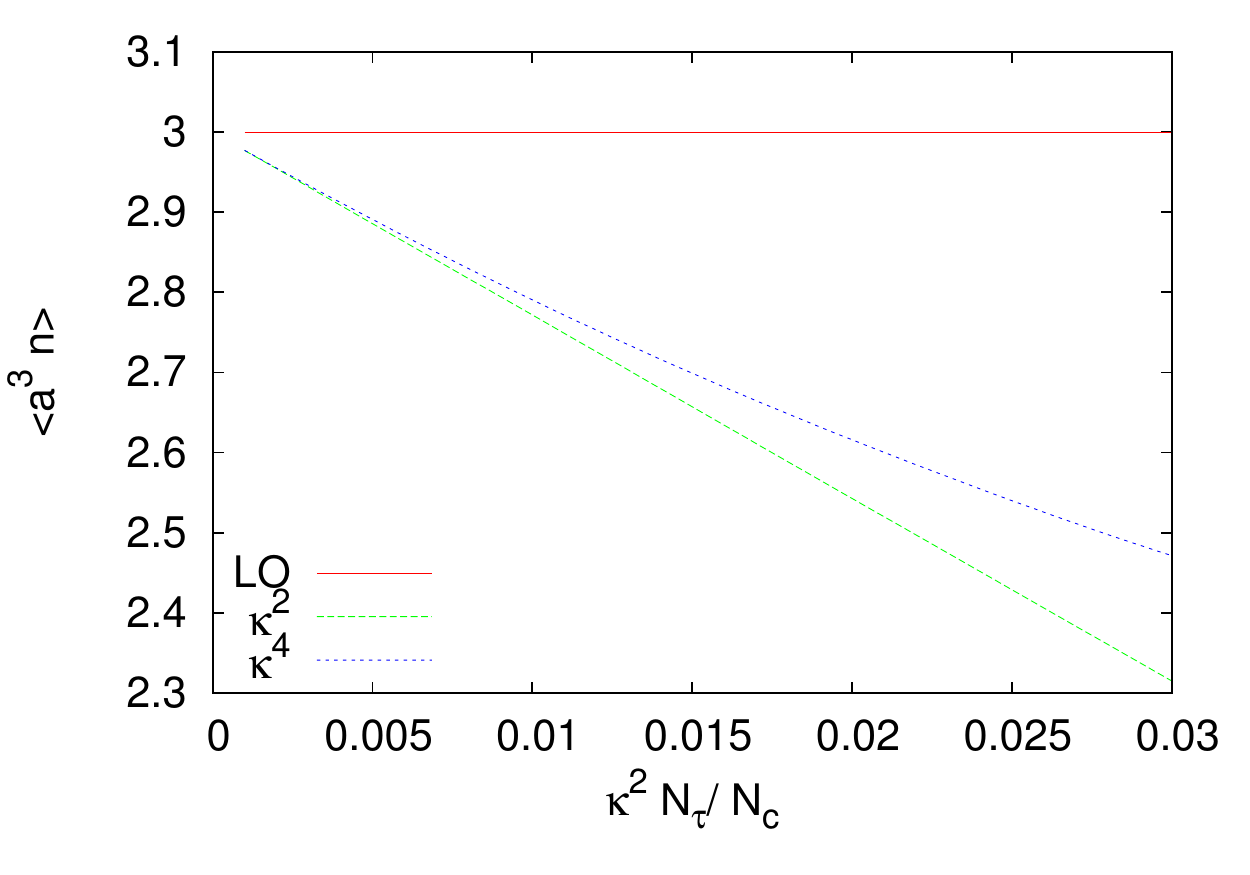}
\includegraphics[width=0.4\textwidth]{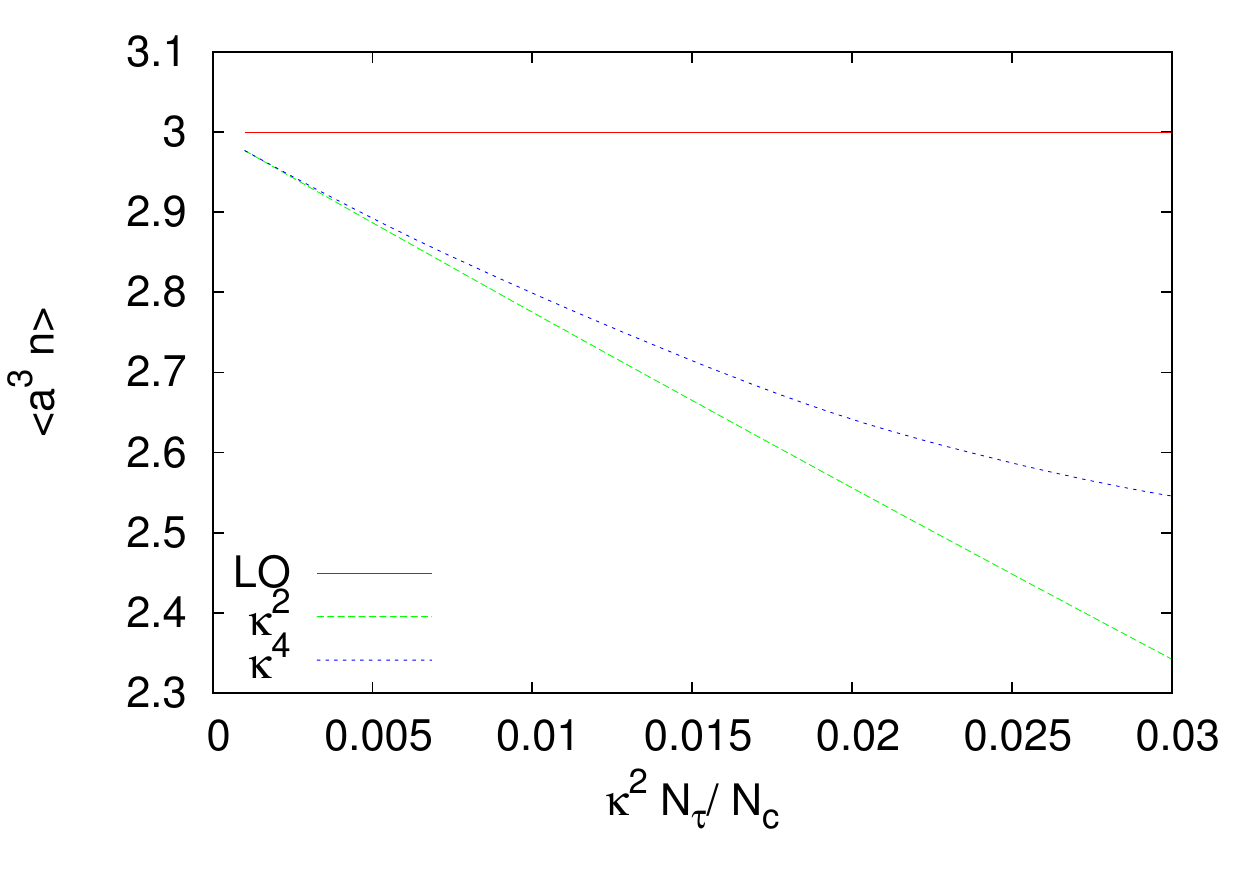}
}
}
\caption[]{Comparison between different orders in $\kappa$, using the standard
action (left), the resummed action (middle) and including gauge corrections
with $\beta=6$ (right).}
\label{fig:convergence}
\end{figure}
The effective theory specified in the last sections has a sign problem. With less
degrees of freedom 
and the theory being only three-dimensional, the sign problem is milder than in the
original theory
such that Monte Carlo methods are feasible at finite temperatures and chemical
potentials
$\mu/T\lsim 3$ \cite{Fromm:2011qi}.
If, however, one is interested in cold dense matter in the zero
temperature limit, the sign problem becomes strong and Monte Carlo methods fail on
large volumes.
Fortunately, the effective theory is amenable to simulations using complex Langevin 
algorithms (for an introductory review, see \cite{dh}) and the onset transition to
nuclear matter could be demonstrated explicitly for 
very heavy quarks \cite{Fromm:2012eb}. In this section we discuss the validity of
this approach for the effective 
theory.  We will only sketch the general method here, as there is an abundant
literature on this subject 
\cite{dh,etiology,su3lang}.

The basic idea is to introduce a fictitious Langevin time $\theta$, in which a field
theoretical 
system with Gaussian noise $\eta(x,\theta)$ evolves according to the Langevin equation
\beq
\label{langevin-eq}
\frac{\partial \phi(x,\theta)}{\partial \theta}=-\frac{\delta S}{\delta
\phi(x,\theta)}+\eta(x,\theta)\;.
\eeq
In the case of a complex action, the field variables have to be complexified too, 
$\phi\rightarrow \phi_r + i\phi_i$. 
In our case, after integration over the spatial links, the degrees of freedom are
the traced 
Polyakov lines
\beq
\int \left[\prod_{\tau=1}^{N_{\tau}} d U_0(\tau) \right] f(L,L^*) = \int d W
f(L,L^*)\;.
\eeq
We may further simplify this by parametrizing the Polyakov lines in terms of two
angles and 
bring them into a diagonal form \cite{gross83}
\beq
L(\theta,\phi) = e^{i \theta}+e^{i \phi}+e^{-i (\theta+\phi)},
\eeq
which introduces a potential term denoted by $e^V$ with
\beq
V=\frac12  \mathrm{ln}(27-18|L|^2+8 \mathrm{Re}(L^3)-|L|^4)\;.
\eeq
Hence the integration measure we use in our simulation is the reduced Haar measure
\beq
\int d W = \int dL e^V = \int_{-\pi}^\pi d\theta \int_{-\pi}^\pi d\phi \ e^{2V}\;.
\eeq
This means instead of an integration over $N_{\tau}$ SU(3) matrices we have 2
complex degrees of freedom on every spatial lattice point. 
Furthermore, having only diagonal matrices their inversion 
is trivial.
With these ingredients eq.(\ref{langevin-eq}) was solved numerically using stepsizes
of around $\epsilon = 10^{-3}$ and applying the adaptive stepsize technique proposed
in \cite{adaptive-stepsize} to avoid numerical instabilities.
\subsection{Criteria for correctness}
Unfortunately, it is well known that the complex Langevin algorithm is not a general
solution to the 
complex action problem as it converges to the wrong limit in some cases, including
some
parameter ranges for QCD \cite{dh,amb86}. The failure can be attributed to
insufficient localisation of
the probability distribution in the complex field space, and a set of criteria was
developed 
to check whether this localisation is sufficient \cite{etiology}. A necessary
condition is that
the expectation value of all observables vanishes after a Langevin operator
$\hat{L}$ has been 
applied to them,
\beq
\langle \hat{L}O[\phi]\rangle=0, \quad
\hat{L}=\sum_{a,x}\left(\frac{\partial}{\partial \phi_a(x)}
-\frac{\partial S}{\partial \phi_a(x)}\right)\frac{\partial}{\partial \phi_a(x)}\;.
\label{langevan operator}
 \eeq
 While, strictly speaking, this test is necessary on {\it all} observables of the
theory, in practice only
 a select few can be tested. In figure \ref{fig:convcrit} we show the expectation
value of the Polyakov loop 
 as a function of the step size of the Langevin algorithm for the effective theory
to order $\kappa^2$ (left)
 and $\kappa^4$ (right). In both cases the criterion is fulfilled.  

As a further and complementary check of the validity of the complex Langevin
simulation, we also
compare with reweighted Monte Carlo results where this is possible, i.e.~on small
volumes. As 
figure \ref{fig:cfMC} shows, this test is also passed by the complex Langevin data
for the 
expectation value
of the Polyakov loop as well as the baryon number density.

\subsection{Convergence region of the hopping series}
\begin{figure}[t]
\centerline{
\includegraphics[width=0.5\textwidth]{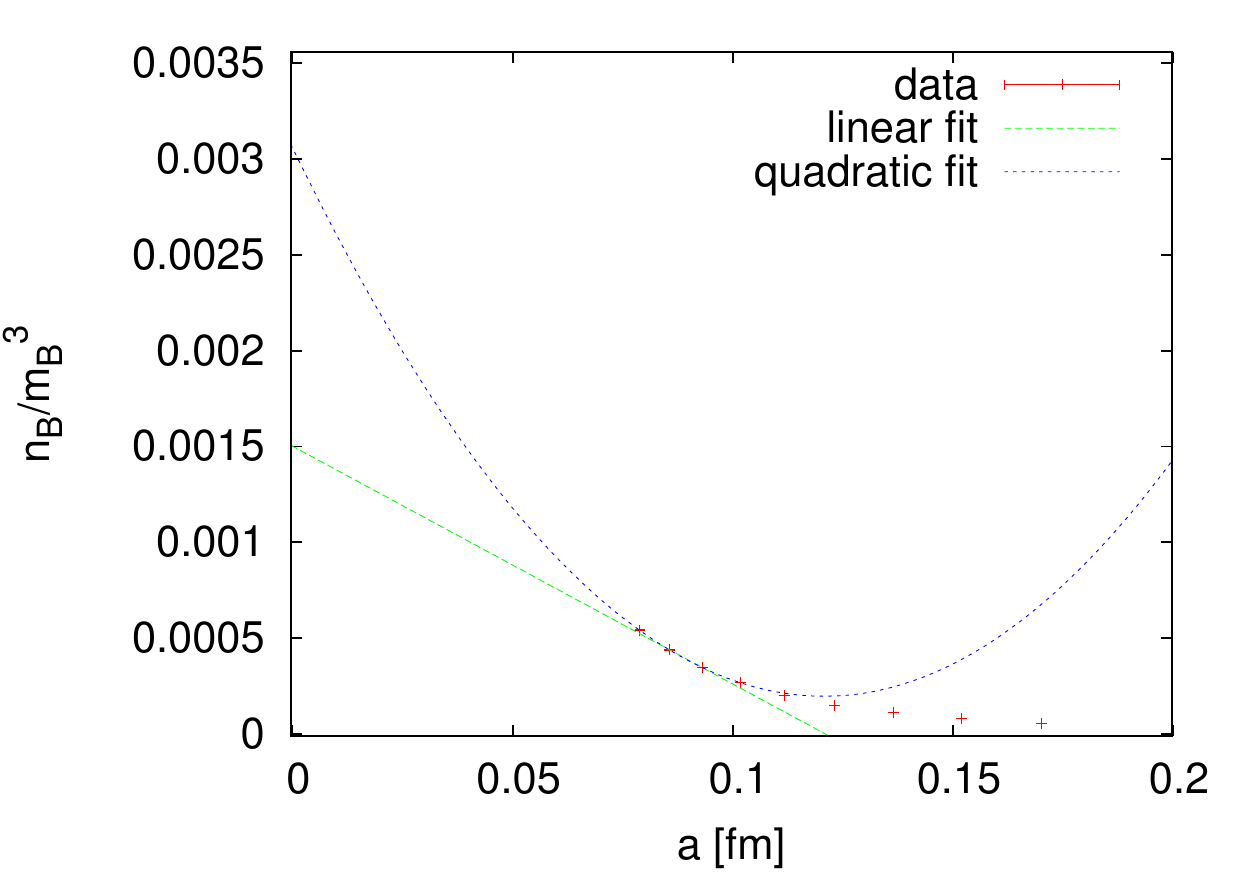}
\includegraphics[width=0.5\textwidth]{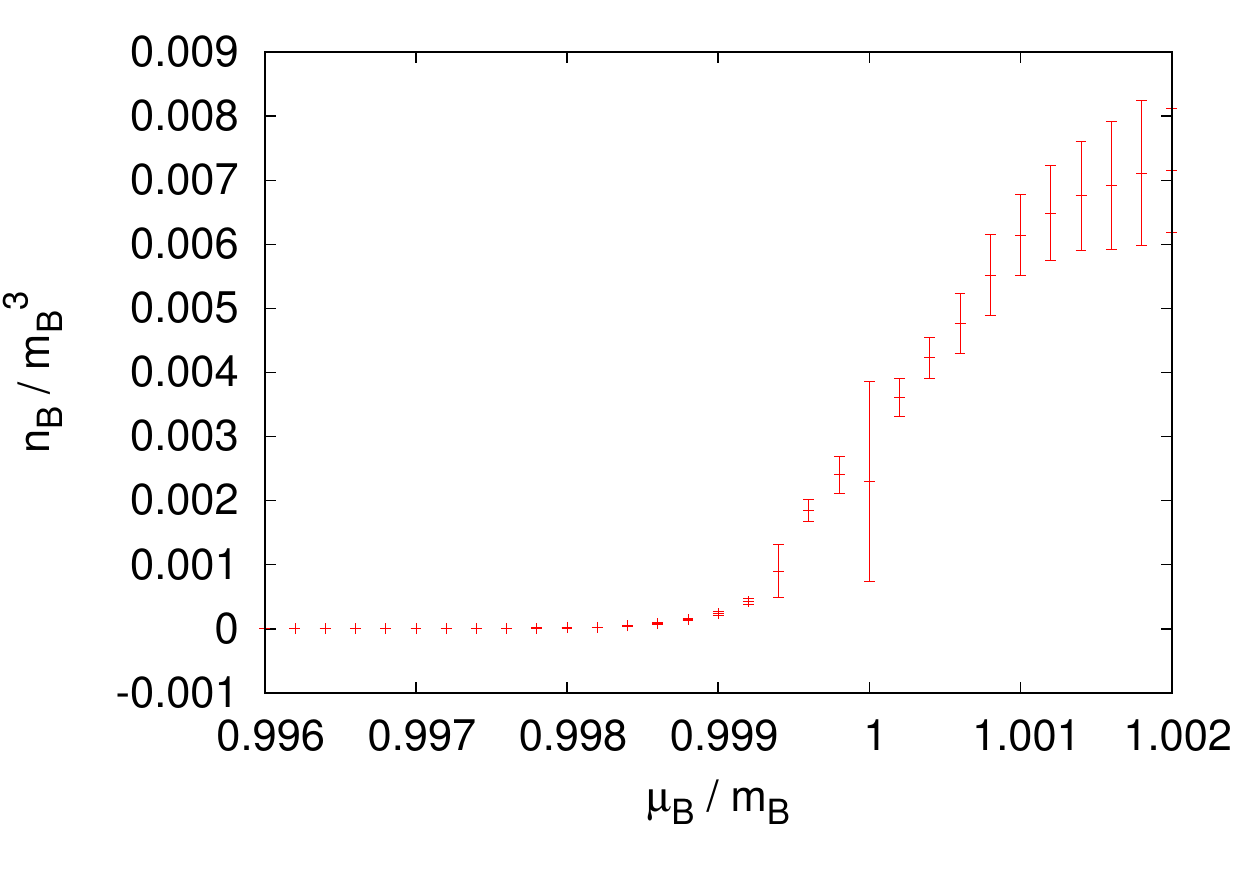}
}
\caption[]{
Examples for the continuum extrapolation. Shown are linear and quadratic extrapolations with one d.o.f. (left)
and continuum extrapolated results for the transition to cold nuclear matter
for $T=10$ MeV (right). In order to set the scale we use results for $r_0$ from \cite{Necco:2001xg} and the strong coupling and hopping parameter expansion of the baryon mass from \cite{Fromm:2012eb}.
 }
\label{fig:silver}
\end{figure}
One of the most important values we want to determine is the region of convergence
of the effective 
theory. This is the region where the truncated theory is a good approximation to the 
full theory.
As criteria for convergence we choose the difference between expectation values
obtained from 
the $\kappa^2$ and the $\kappa^4$ action for different values of the expansion
parameter 
$\frac{\kappa^2 N_{\tau}}{N_c}$. The expansion parameter already shows that the
region of 
convergence is limited in the direction of low temperatures and light quarks,
i.e.~one can 
reach lower quark masses by raising the temperature.
As an observable we choose the density in lattice units $<a^3 n>$ with a chemical potential chosen such that $h_1(\kappa,u,\mu) = 1$.
As can be seen in figure \ref{fig:convergence} the static limit is only a valid
approximation in 
the $\kappa \to 0$ limit. If we define the limit of the region of convergence as a
certain value of 
the difference between $<a^3 n>_{\kappa^2}$ and $<a^3 n>_{\kappa^4}$ one can see
that the resummed 
action offers a better convergence. Therefore, we will use this version for our
simulations.

\subsection{Silver blaze property and onset to nuclear matter}
In our previous work \cite{Fromm:2012eb} we performed a continuum extrapolation for
the transition 
to cold nuclear matter based on the $\kappa^2$ action. 
In figure \ref{fig:convergence} we repeat this calculations including the 
$\kappa^4$ corrections. This allows us to simulate smaller lattice spacings 
$a=0.08$ fm without leaving the 
region of convergence, since reducing $a$ while keeping $\frac{M}{T}$ fixed means
going to higher 
$\kappa$. Nevertheless the 
extrapolation suffers from considerable uncertainties, resulting in large errors in
the high density phase. 
This can be seen in fig. \ref{fig:silver} (left), where we show the two best fits for our data at $\frac{\mu}{m_B}=1$ at several lattice spacings. This is the chemical potential where different extrapolation fits differ the most.
The truncation error for our $\kappa^4$ data is estimated as the difference to the data obtained from the $\kappa^2$ action. This data was  then fitted to get a value for $a \rightarrow 0$. \\
As continuum result we took the average of the two best fits, the error was estimated as difference between those two fits. For each value of the chemical potential we tried several fits (linear and quadratic) with one to three degrees of freedom. For the best fits we always achieved $\chi^2_{red}<2$ as long as $\frac{mu_B}{m_B} < 1.0014$.
The growing uncertainties in the high density region are caused by the unphysical 
saturation on the lattice which limits the density to $2 N_c$ quarks per 
lattice site, while in the continuum no such saturation exists.
In the low density region the Silver Blaze property, i.e. the independence from
chemical potential 
in the $T \to 0$ limit below a critical value $\mu_c$, can be seen. 
Note that the results at $\kappa^4$ are somewhat higher than  
our $\kappa^2$-results in \cite{Fromm:2012eb}. The inclusion of $\kappa^4$ allows
for a better estimate of
the truncation error and therefore inclusion of data from lattices with 
smaller lattice spacing.

\begin{figure}[t]
\hspace*{-2.5cm}
\scalebox{1.3}{
\centerline{
\includegraphics[width=0.25\textwidth]{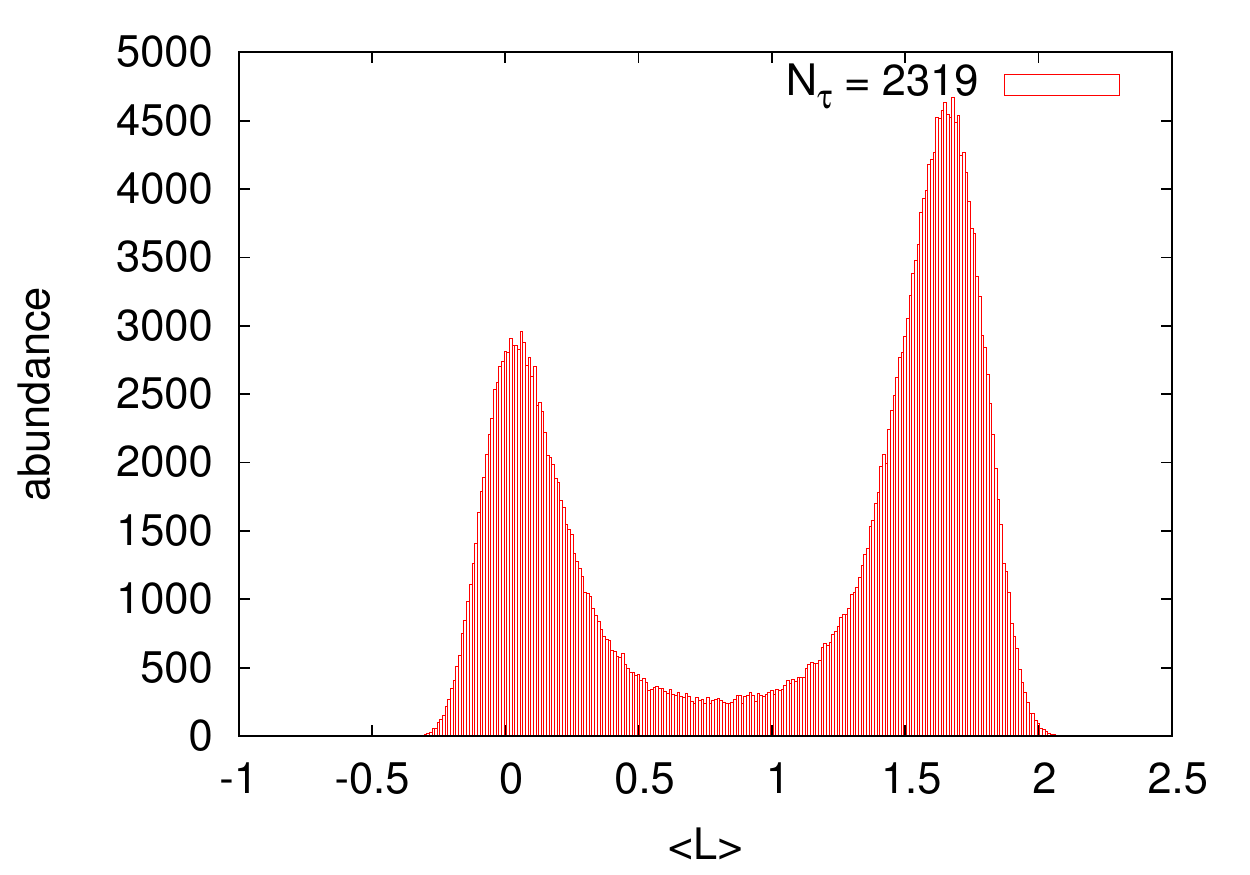}
\includegraphics[width=0.25\textwidth]{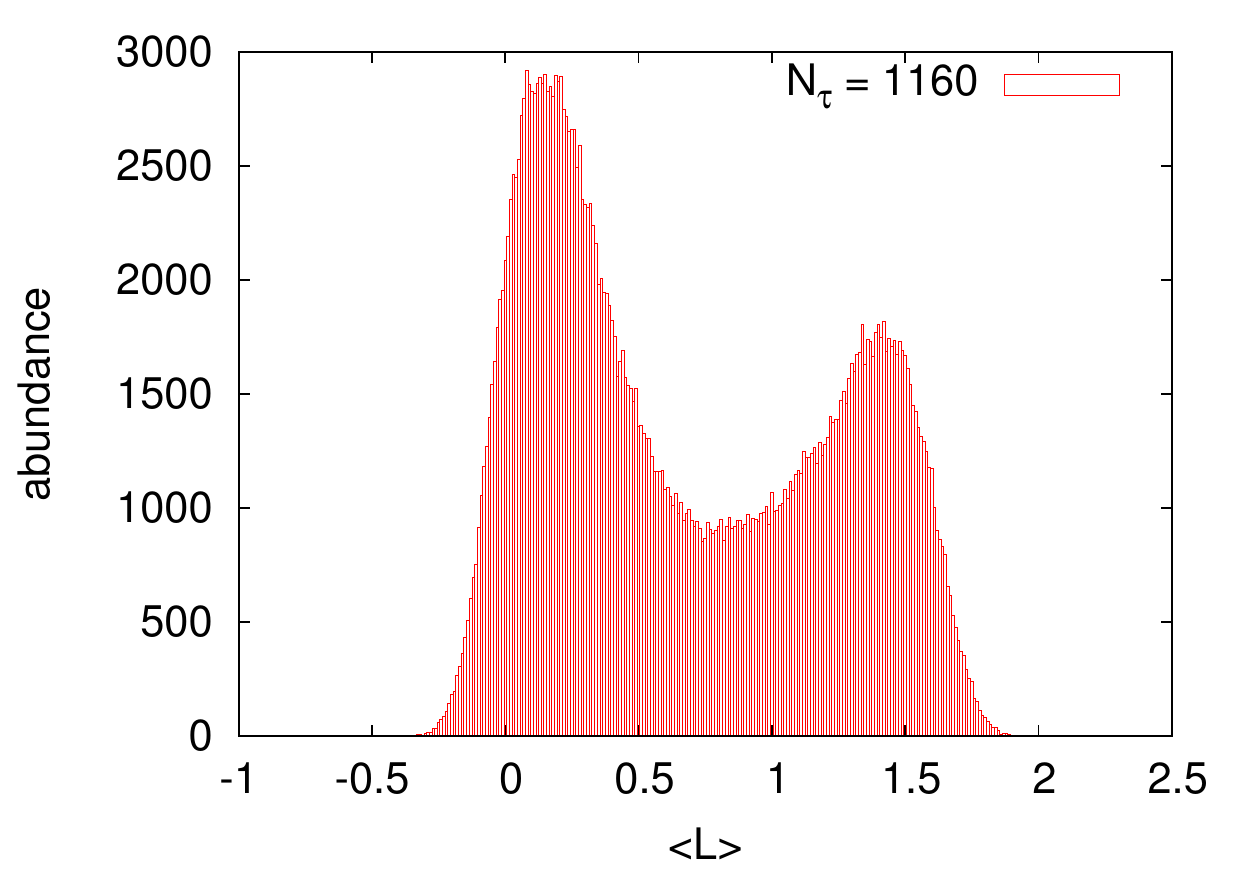}
\includegraphics[width=0.25\textwidth]{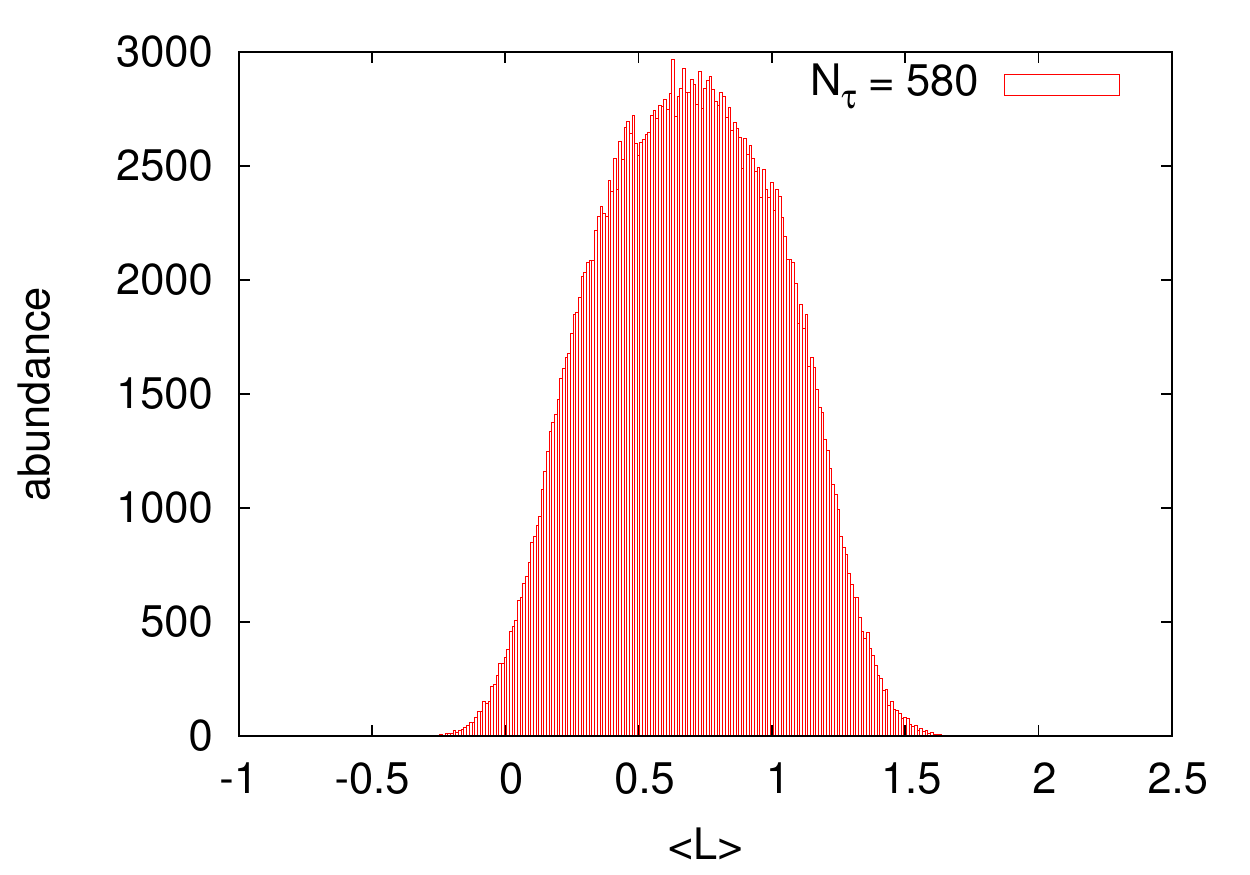}
}
}
\caption[]{Polyakov Loop histograms in the transition region for three different
temperatures, $\kappa = 0.12$ and $\beta = 5.7$. }
\label{fig:polyakov-hist}
\end{figure}
As in our previous work \cite{Fromm:2012eb}, the accessible quark masses in
the convergence region of the effective theory are 
too high to realize the expected first order transition from the vacuum to the 
region of finite density, i.e.~the transition proceeds as a smooth crossover.
However, this changes if we leave the convergence region by going to lower quark
masses 
($\kappa = 0.12$) and very 
low temperatures with $N_{\tau} = O(10^3)$, where 
we see signals for a first order 
transition. Figure \ref{fig:polyakov-hist} shows distributions of Polyakov Loop
expectation values 
in the transition region. It clearly shows coexistence of two distinct phases, 
i.e.~the
effective action describes a first order transition which disappears as 
temperature is raised ($N_\tau$ is lowered). However, in order to 
make quantitative statements 
we will have to extend the region of convergence by adding several 
higher orders in $\kappa$.

\section*{Acknowledgements}
J.L. is supported by the Swiss National Science Foundation under 
grant 200020-137920. M.N. and O.P. are  supported by the German BMBF, 
grant 06FY7100, and the Helmholtz International
Center for FAIR within the LOEWE program launched by the State of Hesse.

\end{document}